\documentstyle[12pt]{article}
\input math_macros.tex

\def\ref#1{$^{#1)}$}
\begin{document}
\begin{titlepage}
\begin{center}
\today  
  \hfill LBNL-42649  \\

\vskip .5in

\centerline{\large\bf}

\vskip .3cm
\centerline{\large\bf Elementary-Particle Propagation Via 3-Scale}
\vskip .3cm

\centerline{\large\bf ``Towers of Quartet Rings'' Within a Dyonic History Lattice}

\vskip .5in

G.F. Chew\\[.5in]

{\em Theoretical Physics Group\\
     Physics Division\\
     Lawrence Berkeley National Laboratory\\
     Berkeley, California 94720}
\end{center}

\vskip .5in

\end{titlepage}
\renewcommand{\thepage}{\roman{page}}
\setcounter{page}{2}
\mbox{ }

\vskip 1in

\begin{center}
{\bf Disclaimer}
\end{center}

\vskip .2in

\hsize=6in
\baselineskip=14pt

\begin{scriptsize}
\begin{quotation}

This document was prepared as an account of work sponsored by the United States 
Government.  Neither the United States Government nor any agency thereof, nor
The  Regents of the University of California, nor any of their employees, makes
any warranty,  express or implied, or assumes any legal liability or
responsibility for the accuracy,  completeness, or usefulness of any
information, apparatus, product, or process disclosed,  or represents that its
use would not infringe privately owned rights.  Reference herein to  any
specific commercial products process, or service by its trade name, trademark, 
manufacturer, or otherwise, does not necessarily constitute or imply its
endorsement,  recommendation, or favoring by the United States Government or
any agency thereof, or  The Regents of the University of California.  The views
and opinions of authors expressed  herein do not necessarily state or reflect
those of the United States Government or any  agency thereof of The Regents of
the University of California and shall not be used for  advertising or product
endorsement purposes. \end{quotation} \end{scriptsize}

\vskip 2in

\begin{center}
\begin{small}
{\it Lawrence Berkeley National Laboratory is an equal opportunity employer.}
\end{small}
\end{center}

\newpage
\renewcommand{\thepage}{\arabic{page}}
\setcounter{page}{1}

                \hfill    \today  \\
\vskip1.3cm

\vskip .3cm
\centerline{\large\bf Elementary-Particle Propagation Via 3-Scale}
\vskip .3cm

\centerline{\large\bf ``Towers of Quartet Rings'' Within a Dyonic History Lattice}
\vskip .6cm
\centerline{\bf G. F. Chew}
\vskip .3cm

\centerline{Theoretical Physics Group, Lawrence Berkeley National  Laboratory}
\centerline{One Cyclotron Road}
\centerline{Berkeley, California 94720, U.S.A.}
\vskip 1.1cm

\begin{quotation}

\tenrm\baselineskip =  12pt
\centerline{\bf ABSTRACT}
\vskip .2cm

Massless elementary-particle propagation is represented historically
 (cosmologically) through 3-scale ``towers of quartet rings''  within 
a lattice of magneto-electrodynamically communicating ``pre-events''.  
The lightlike intervals within a ring of 4 pre-events (discrete 
``closed string'') display transverse GUT-scale and longitudinal 
``particle scale''.  The lightlike (longitudinal) spacing between 
successive rings of a tower is at Planck scale.  Ratio between GUT 
scale and Planck scale relates quantum-dynamically to elementary 
magnetic charge.  Permutations of a ring quartet, in conjunction with 
Lorentz-group representations, control elementary-particle quantum numbers.

\end{quotation}
\vskip 1cm

\noindent
{\bf I.\ \ Introduction}
\vskip .3cm

A historical quantum cosmology, based on lattice coherent states
labeled locally by ``pre-event'' chains, has been formulated$^{(1)}$.  
The constraints defining ``history lattice'' include causal (i.e.,
lightlike) magneto-electrodynamic (MED) pre-event connections. $^{(2)}$  The present paper
offers 
historical representation of massless elementary-particle propagation 
through a special history pattern describable as a ``tower of
magnetically-
confined pre-event quartet rings'' that display both GUT scale and 
a much larger ``particle scale''.  Spacing between successive rings 
is at a third, more fundamental, local scale that historical cosmology
identifies
with Planck scale\footnote {This paper ignores a
  global scale, hugely larger than any of the 3 scales discussed here
  (although smaller than local age), that relates to distinction
  between ``present'' and ``past''$^{(1)}$.}.  Ratio between GUT and Planck scales will here be 
related to elementary magnetic charge.\footnote {Magnetic charge is 
screened at scales above those characterizing quartet rings  while 
localized energy, defined through towers, is devoid of meaning 
at lower scales.  Historical cosmology thus precludes ``magnetic
charge 
carried by matter.''}  

Although the present paper does not address 
historical representation of interaction $\it {between}$ elementary 
particles, many of the principles invoked here are relevant to
particle-
interaction (e.g., ``pair of pants'') pre-event patterns -- 
which may be called ``local events'' in contrast to ``pre-events.''  
Our concluding section calls attention to one such principle: the 
continuity, along a history-chain segment, of electric charge and 
certain other discrete ``quantum numbers''.

A tower-propagator history pattern might be described as a ``closed
string moving in spacetime'', but the ``string'' is discrete 
(comprising 4 pre-events with GUT-scale spatial spacing and 
particle-scale temporal spacing) and its ``motion'' is discrete 
(in Planck-scale steps).  Despite history-lattice discreteness, 
the (continuous) Lorentz group plays a central historical role, 
discreteness being reserved to Lorentz invariants.  Local physics 
blends into global cosmology within a discrete history that
relies on the Lorentz group at all scales.   Local Lorentz
covariance, while not an exact feature of historical cosmology, is 
approximately valid for velocities currently accessible to physical 
observation.
 
\vskip .4cm

\noindent
{\bf II.\ \ Towers of Interleaved Pre-event Ring Quartets} 
\vskip .3cm

Reference (1) prescribes a ``history lattice'' with a longitudinal
``chain'' structure.  Matter representation requires further a transverse ``ring'' structure.  
Steps along the chain and intervals around a ring are both lightlike.  
The longitudinal chain step has a $\it{fixed}$ interval $\delta$ in
pre-event age (``age'' is defined below).  It has been conjectured 
that the value of $\delta$, once meaning for the gravitational 
constant has been identified within historical cosmology, will 
turn out to be on Planck scale.  The fundamental parameter $\delta$ 
establishes a ``smallest'' scale for historical cosmology.  

Although longitudinal chains comprise Hubble-scale closed loops, 
to the extent that each longitudinal loop contains a number of
pre-events ($>10^{60}$ for loops passing close to our region of the
universe) much larger than the pre-event number building a typical 
propagator tower (particle collision times are presumed small on 
Hubble scale), this paper will disregard the compactness of 
(longitudinal) history-chain segments.  Such a posture conforms 
to physics disregard of distinction between ``present'' and ``past''.  

The number of pre-events in a ``transverse ring'' is exactly 4 with 
each member of a pre-event ring quartet belonging to a $\it
{different}$ segment of the (longitudinally-looping) history chain.  
The term ``ring'' means the 4 pre-events are cyclically ordered.  
$\it {Adjacent}$ pre-events around a ring have $\it {opposite}-{\it sign}$ 
magnetic charge and lightlike spacetime separation (they
``communicate'', with alternation of advanced and retarded connections).
In the ring quartets building propagator towers, the magnitude of any 
pre-event magnetic charge is the elemental $\tilde{e}$; passing around a
ring, magnetic-charge sign alternates.  ``Opposing'' pre-events 
(``across the ring'' from each other) share not only the same magnetic 
charge (both +$\tilde{e}$ or both $-\tilde{e}$) but the same age.  Opposing 
pre-events do not (directly) communicate with each other, being
spacelike separated.  The mean spatial separation between opposing 
pre-events will be found to exhibit GUT scale. 

The magnitude of the age difference (an integer multiple of $\delta$)
between magnetically-positive and magnetically-negative pre-events
within a quartet is common to $\it {all}$ rings within tower
(matter-propagation) history patterns.  This difference will be
designated by the symbol $\delta'$.  Section III will associate
the term ``particle scale'' with  $\delta'$ and argue that $\delta'$ 
must be huge on GUT scale in order to achieve the approximate local
Lorentz covariance of particle physics.  A matter-representing ring
quartet exhibits GUT scale spatially and particle scale temporally, 
with Planck scale characterizing 
the spacing between successive rings building a tower.  

According to Reference (1), $\it{all}$ history (whether tower or
otherwise) occupies the interior of a forward big-bang lightcone.  
Such an idea first appeared in the cosmology of Milne.$^{(5)}$ 
Spacetime location of Pre-event n is given by a positive-timelike
4-vector 
$\it {\bf x}_n$ prescribing pre-event $\it {displacement}$ from lightcone
vertex.  
Pre-event age $\tau_{n}$ is the Lorentz magnitude of ${\bf x}_n$.  
(Pre-events in $\it{our}$ neighborhood have ages of order $10^{10}$ 
years or $\sim 10^{60}$ in units of $\delta$).

Let us designate those 4 pre-events  building a ring quartet 
Q by the index pair (Q, i) with i cyclic -- i.e., i${\pm}$4 is the
same as i.  A pre-event pair (Q, i) and (Q, i $\pm$ 1) are ``adjacent''; 
their lightlike separation means

$$
( {{\bf x}_i}^Q - {\bf x}^{Q}_{i\pm1})^2 = 0, i = 1, 2, 3,
    4.\quad \eqno {\rm (1)}$$ \\ A pair $(Q, i)$ and  $(Q, i\pm2)$ are ``opposing'', with spacelike
separation:

$$( {{\bf x}_i}^Q - {\bf x}^{Q}_{i \pm2})^2 = -{({d}_{-}^Q)}^2,
i = 1, 3 \quad ,\eqno {\rm (2)}$$

$${({\bf x}_i^Q - {\bf x}_{i\pm2}^Q)}^2 = -{(d_+^Q)}^2, i = 2, 4.\quad
\eqno {\rm (3)}$$ \\ The notation in (2) and (3) reflects assignment of negative magnetic
charge to i = 1, 3 and positive magnetic charge to i = 2,
4.\footnote{Electric charge and other discrete quantum numbers render
  distinct all 4 values of i.}  The (length dimension) spatial parameters ${d}_{\pm}^Q$ prescribe the 
``shape'' and ``size'' of the ring.

The set of four 4-vectors ${\bf x}_i^Q$, comprising 16 parameters,
satisfy the following conditions on the 10 Lorentz invariants,
${\bf x}_i^Q \cdot {\bf x}_j^Q$:

\vskip .4cm

$$ {\bf x}_{1}^{Q} \cdot {{\bf x}_{1}}^Q = 
 {{\bf x}_{3}}^Q \cdot {\bf x}_{3}^Q =
(\tau_{-}^{Q})^{2}\quad ,\eqno {\rm (4)}$$

$$ {\bf x}_{2}^{Q} \cdot {{\bf x}_{2}}^Q =
 {{\bf x}_{4}}^Q \cdot {\bf x}_{4}^Q =
(\tau_{+}^{Q})^{2}\quad ,\eqno {\rm (5)}$$

$$ {\bf x}_{1}^{Q} \cdot {{\bf x}_{3}}^Q =
 (\tau_{-}^{Q})^{2} + 
\frac{{(d_-^Q)}^{2}}2\quad ,\eqno {\rm (6)}$$

$$ {\bf x}_{2}^{Q} \cdot {{\bf x}_{4}}^Q =
 (\tau_{+}^{Q})^{2} 
+ \frac{{(d_+^Q)}^{2}}2\quad ,\eqno {\rm (7)}$$

$$ {\bf x}_{i}^{Q} \cdot {{\bf x}^{Q}_{i\pm1}} =  \frac
{{(\tau_+^Q)}^2 + {(\tau_-^Q)}^2}{2}, \> \hspace{.1in}i = 1, 2, 3, 
4.\quad
\eqno {\rm (8)}$$ \\ Thus all 10 invariants are determined by the 4 positive parameters,
${d_{\pm}}^Q, \tau_{\pm}^Q$.  According to Section III below, the length
parameters $d_{\pm}^Q$ are not fundamental constants and relate
quantum-magneto-dynamically to the quartet of pre-event impulses.

There remain six ``external'' ring degrees of freedom -- the parameters
of a Lorentz transformation that leaves invariant the 10 inner
products ${{\bf x}_i}^Q \cdot  {{\bf x}_j}^Q$.  Defining the ring-center
(positive-timelike) 4-vector

$${\bf X}^Q \equiv \frac {1}{4}({\bf x}_1^Q + {\bf x}^Q_2 + {\bf
  x}^Q_3 + {\bf x}^Q_4), \quad
\eqno {\rm (9)}$$ \\ with magnitude (center age)

$${\tau^Q \equiv \{{\frac {1}{2}[{(\tau^Q_+)}^2 + {(\tau^Q_-)}^2] + \frac
  {1}{16}[{(d^Q_+)}^2 + {(d^Q_-)}^2]\}^{\frac {1}{2}}}}, \quad \eqno 
{\rm (10)}$$ \\ 3 Lorentz boost parameters spatially locate ring center with respect
to big-bang vertex,
\footnote{Designating the 3 ring-center boost
  parameters by a 3-vector $\vec {\beta}^Q$, spatial location of ring
  center is \\
$${\vec {X}}^Q = \tau^{Q} \frac{ \vec {\beta}^Q}{|\vec {\beta}^Q|} sinh
| \vec {\beta}^Q|.$$}
while 3
rotation parameters orient the ring in
its center Lorentz frame.  \\Counting center age, a ring has a total of 7 ``external'' degrees of
freedom.  (When $\it different$ rings combine into a larger history pattern,
such as a tower, some of the separate-ring external degrees of freedom
become ``internal'' to the larger pattern.  The {\it total} number of
external degrees of freedom for {\it any} pattern is 7, corresponding 
to a Lorenz transformation plus an age.)

The coordinates of any quartet provide a complete basis.  Employing the notation 

$${\bf x}_{ij}^{Q} \equiv {\bf x}_i^Q - {\bf x}_j^Q, \quad \eqno 
{\rm (11)}$$ \\ a convenient basis comprises the two spacelike 4-vectors ${\bf
 x}_{24}^Q, {\bf x}_{31}^Q$  and the two timelike 4-vectors ${\bf X}^Q$ and 

$${\bf U}^Q \equiv \frac{1}{2} ({\bf x}_2^Q + {\bf x}_4^{Q}) - \frac
{1}{2} ({\bf x}_1^Q + {\bf x}_3^Q), \quad \eqno {\rm (12)}$$ \\ with magnitude

$$d^Q = \frac{1}{2} {[{(d_+^Q)}^2 + {(d_-^Q)}^2]}^{\frac{1}{2}}. \quad
  \eqno {\rm (13)}$$ \\ We may render
${\bf U}^{Q} \it positive$-timelike by requiring
$\tau_{+}^{Q} > \tau_{-}^{Q}$.

The spacelike pair ${\bf x}_{24}^Q$, ${\bf x}_{31}^Q$ are orthogonal not only to
each other and to ${\bf U}^Q$ but to ${\bf X}^Q$.  Only the timelike
basis pair ${\bf X}^Q, {\bf U}^{Q}$ fails to be orthogonal.  The two spacelike directions
may be called ``transverse'' and the two timelike directions
``longitudinal''. \\ In the proposed basis, 

$${\bf x}^Q_1 = {\bf X}^Q - \frac{1}{2} {\bf U}^Q - \frac {1}{2} {\bf
  x}^Q_{31},  \quad \eqno$$

$${\bf x}^Q_2 = {\bf X}^Q + \frac{1}{2} {\bf U}^Q + \frac {1}{2} {\bf
  x}^Q_{24}, \quad \eqno {\rm (14)}$$

$${\bf x}^Q_3 = {\bf X}^Q - \frac{1}{2} {\bf U}^Q + \frac {1}{2} {\bf
  x}^Q_{31}, $$

$${\bf x}^Q_4 = {\bf X}^Q + \frac{1}{2} {\bf U}^Q - \frac {1}{2} {\bf
  x}^Q_{24}.  \quad \eqno {\rm}$$ \\ The reader may find it a useful exercise to verify the relations (1),
(2), (3), through (14) given (13).  For subsequent use we also record
here the relation

$${\bf X}^{Q} \cdot {\bf U}^{Q} = \frac{1}{2}[{(\tau_{+}^{Q})}^2 -
{(\tau_{-}^{Q})}^2] + \frac{1}{8}[{(d_{+}^{Q})}^2 - {(d_{-}^{Q})}^2],
\quad \eqno {\rm (15)}$$
that is deducible from formulas (4).....(8).  

There is a lightlike displacement from Pre-event (Q, i) to the
following (older, by the increment $\delta$) pre-event along its history-chain
segment.  A unique set of 4 displacements, connecting a pair of
longitudinally-adjacent quartets within a tower, is determined by
requiring that quartet orientation and ``shape'' remain unchanged
throughout a tower, and that displacement of quartet center be
orthogonal to the two (fixed) orthogonal transverse spacelike
directions.  By ``shape'' is meant the ${\it ratio}$ of the two
transverse distances $d_+^Q/d_-^Q$.  A tower is thus characterized by the
orientation and shape of ${\it any}$ of its constituent ring quartets.
The direction of tower ``axis'' corresponds to propagation
direction.  The magnitude $d^{Q}$ varies from one quartet to the next as
the age of each pre-event increases by the increment $\delta$,
but in a ``mature'' (``dilute'')region of the universe,
where $\tau
  \gg {(\delta'/\delta)}^2 \delta$, the fractional variation is of
  order $\delta/\tau$.  
For purposes of
mature-region
``physics'' (as opposed to ``cosmology''), a tower is characterized by
a $\it single$ ``width'' d.  Quantum fluctuation of d is nevertheless
important.

Notice that, with $\delta' > \delta$, age spacing between
magnetically-positive and magnetically-negative pre-events within the
same ring quartet is larger than spacing between same-sign pre-events
of (longitudinally) adjacent quartets.  There is ${\it interleaving}$ of
ring quartets within a tower.  The electrodynamic considerations of
the following section indicate a value of $\delta'$ huge
compared to $\delta$.

Collision times (age intervals between successive collisions of
elementary particles) are meaningless if smaller than $\delta'$ because
individual ring quartets are unsustainable.  More generally, the tower
history pattern is essential to the very meaning of ``matter''.
Within historical cosmology, matter (localized energy) is only an
approximate concept, meaningful at spatial scales larger than d and
temporal scales larger than $\delta'$.  Extremely close to big
bang where towers are unsustainable, localized energy loses
significance.

\vskip .4cm

\noindent

{\bf III.\ \ Pre-Event Impulses Within a Tower }

\vskip .3cm

Each pre-event is endowed not only with a positive-timelike
spacetime-location ${\bf x}_n$ but with a spacelike 4-vector impulse
${\bf q}_n$
and a positive timelike unit-magnitude charge 4-velocity ${\bf u}_n$ that
is orthogonal to $\it {\bf q}_n$.  In a tower, as this section will
elaborate on the basis of Reference (2), charge velocities are
longitudinal while impulses are transverse.  Reference (2) prescribes
history-lattice constraints that ``causally'' (and linearly) determine
the impulse at a pre-event from the electric and magnetic charges and
charge velocities at those {\it other} pre-events located on its
lightcone.  (Impulses, like spatial separations between pre-events, are
subject to quantum fluctuation.)

Within a tower history pattern in a mature region of the universe,
there is near MED cancellation from ``sources'' on that history-chain
segment containing the pre-event whose impulse is being calculated.
We concentrate here on MED interactions within a
{\it single} ring quartet, there being within a tower {\it no}
lightlike-separated (i.e., ``communicating'') pre-event pairs that
locate on different history-chain segments and {\it also} locate in
different ring quartets.  The interacting pre-event pairs within a
ring quartet are those 4 pairs adjacent to each other around the
ring.  The impulse at any pre-event is generated by the charges of its
two ``ring neighbors'', whose magnetic charge is of sign opposite to
its own.  Electric charge is less constrained.  Each of the 4
history-chain segments may carry electric charge 0, $\pm$ e.  (In
Section IV, photon coupling to electric charge indicates that at least
one of the 4 chain segments building a charged-particle tower carries zero electric
charge.)

In a mature region of the universe, the displacement between centers
of successive quartets is nearly at maximum velocity -- corresponding to
massless-particle propagation.  The success of Lorentz-covariant local field theories (such as
QED) that make no distinction between charge and matter velocity
(and no reference to a special frame) implies a tower charge velocity close to maximum; direction should be
longitudinal.  The unique longitudinal 4-direction ``internal'' to a
quartet is that of ${\bf U}^Q$.  In the absence of alternative we
postulate for any charge velocity within quartet Q, 

$${\bf u }^{Q} = {\bf U}^Q/d^Q. \quad \eqno {\rm (16)}$$ \\ Such
assignment implies spatial direction of $\it {\bf U}^Q$ in the
direction of propagation (${\it not}$ in the spatially reversed
diretion).  

What is meant by a magnitude of charge velocity ``close to maximum''?
Cosmology provides an answer to this question through the special
Lorentz frame belonging to quartet center.  This frame, approximately
shared by all history within a region small on the scale of $\tau^{Q}$
(``Hubble scale''), is the frame in which cosmic background radiation
is approximately isotropic (analog of co-moving coordinates in
standard cosmology).

In quartet-center frame,

$${\bf X}^Q = (\tau^Q, \vec {0}), \quad \eqno {\rm (17)}$$

$${\bf U}^{Q} = d^{Q} (cosh \xi^{Q}, \vec{n}^{T} sinh \xi^{Q}),\quad
\eqno {\rm (18)}$$ \\ where $\vec n^T$ is a unit 3-vector prescribing propagation direction
and 

$$cosh \xi^Q = {\bf X}^Q \cdot {\bf U}^Q/d^Q \tau^{Q}. \quad \eqno
{\rm (19)}$$ \\ Referring to (15), we find in mature-region approximation, 

$${cosh \xi \approx {\frac {\delta'}{d}}}. \quad \eqno {\rm (19')}$$

In center frame, magnitude of charge velocity is 

$${tanh {\xi} \approx  [ 1 - {(\frac {d}{\delta'})}^2]^{1/2}}. \quad \eqno
{\rm (20)}$$ \\ ``Close to maximum'' thus means huge $ \frac {\delta'}{d}$.  The success of
Lorentz-covariant models that tie charge velocity to matter velocity (and ignore the
special Lorentz frame) requires a huge order of magnitude
 for $\delta'/d$,
with the age interval $\delta'$ establishing a scale that we optimistically call
``particle scale'', expecting eventually to connect $\delta'$
with particle-physics data.  Hugeness of cosh $\xi^Q$ means that {\it
  all} longitudinal 4-vectors characterizing {\it internal} tower
structure are ``almost lightlike''; approximate Lorentz covariance for
earth-based particle physics then becomes possible.\footnote {Qualitative successes of
  primordial nucleosynthesis calculations$^{(3)}$ indicate that collision
 times remain meaningful at scales as low as $10^{-23}$ sec.
An ${\it upper}$ limit on $\delta'/\delta$ in the neighborhood of $10^{20}$
is thereby set.  Phenomenological considerations by Coleman and Glashow$^{(6)}$ 
indicate a {\it lower} limit for cosh $\xi$ in the neighborhood of $10^{11}$.}

With $\tilde{e} \gg e$, the largest contribution to the MED impulse at
Pre-event \\  (Q, i) arises from its magnetic charge interacting with
the magnetic charges of its two neighbors.  The prescription of Reference (2)
gives, for magnetic impulse, 

$$q^{Q,mag}_{i, \mu} =
\frac{{\delta \tilde{e}}^2}{{(d^Q)}^3} {(x^Q_{i, i\pm2})}_{\mu},
\quad \eqno {\rm (21)}$$ \\ the direction being ``inwardly-radial'' -- {\it toward} tower central 
axis. (Note that all previous 4-vector relations in this paper have
implicitly involved only ${\it upper-index}$ 4-vectors.)  Radial
magnetic-impulse magnitude is 

$$q_{\pm}^{Q, mag} = \frac{\tilde{e}^2 \delta d_{\pm}^Q}{(d^Q)^3}. \quad
\eqno {\rm (22)}$$ \\ Magnetic ``coulomb attraction'' provides transverse stability for a
tower of quartet rings \footnote {Longitudinal stability is provided
    by persistence of electric and magnetic charges, along with other
    discrete indices, along any segment of the (looping) history
    chain.}.

Electric impulses proportional to $e_i e_{i\pm 1,}$ being smaller than (22)
by a factor of order $(e/\tilde{e})^2$, cannot undo tower stability even
when ``repulsive'' (i.e., when outwardly-radial).  There are no
impulses proportional to $e\tilde{e}$, a product we expect to be
constrained in a manner paralleling that discovered by Dirac$^{(4)}$, once
quantum superposition of history is systematically addressed.  This
paper will not be systematic in such regard.

Formula (22) for the radial impulse, nevertheless, in conjunction with
qualitative quantum considerations, suggests the order of magnitude of
$d^{Q}$.  When towers of different $d_{\pm}^{Q}$ are superposed, minimum
fluctuation (``ground state'')\footnote{Such reasoning, applied to
  the ground state of hydrogen, correctly estimates the ratio of radius
  to period as $\sim {(3e)}^2 /\hbar$.}  is plausibly achieved by
$d_+^{Q}
= d_{-}^{Q}$, with

$$q_{\pm}^Q d_{\pm}^Q \sim \hbar. \quad \eqno {\rm (23)}$$ \\ Formula (22) then yields the estimate

$$<d^{Q}> \sim \frac{\tilde{e}^{2}}{\hbar} \delta \quad \eqno {\rm (24)}$$
\\ for ``mean tower width''.  

Tower structure has now been characterized by 3 different scales --
$\delta$, $\delta'$ and \\ $<d^{Q}>$.  Largeness of $\tilde{e}^{2}/\hbar$
(assuming small fine-structure constant, with $e \tilde{e} \sim \hbar$)
suggests association of mean $d^Q$ with GUT scale.  Propagator-towers
within the history lattice, that is to say, have GUT scale mean
transverse dimension \footnote {Alternatively one might say that
  ``closed strings'' of 4 pre-events have GUT-scale ``radius''.  The
  dynamical considerations of the present section might be regarded as
  an estimate of ``string tension.''}.  A mean value of cosh $\xi$ in
the neighborhood of $10^{16}$ might then be compatible with
$\delta'/\delta \sim 10^{18}$.

\vskip .4cm
\noindent
{\bf V.\ \ Conclusion} 
\vskip .3cm

Although not pursued here, quantum superposition of ${\it external}$
tower Lorentz parameters provides meaning for massless
elementary-particle momentum and helicity via representations of the
Lorentz group\footnote{Fluctuation in transverse location of tower
  center is not to be confused with the tower ``width'' d.}.  Choice of
  representation involves an 8-valued ``sector index'' (in
  addition to electric and magnetic charge) that is carried by each
  history-chain segment.  Four of the 8 values, denoted in Reference
  (1) by $S_1$, $\bar{S}_1$, $S_2$, $\bar{S}_2$,  are suitable for patterns of
  material history.  The index interchanges  $S_1 \leftrightarrow
  \bar{S}_1$ and  $S_2 \leftrightarrow
  \bar{S}_2$  have a  significance paralleling (local) ``particle-antiparticle
  conjugation.''  The 1 $\leftrightarrow 2$ interchange has a global
  significance. 
  The outlook for historical cosmology
  depends on whether ring-quartet patterns of electric charge, spin
  and internal quantum numbers conform to the discoveries of particle physics.
  Permutation-group $(S_4)$ analysis in
  conjunction with Lorentz-group representation must be undertaken.  

The 4-remaining values of sector index, denoted $T_1, \bar{T}_1, T_2,
\bar{T}_2$ in Reference (1), contribute to ``nonmaterial'' patterns of
history that tentatively have been called ``vacuum''.  Coupling of
``vacuum'' to material history is conjectured to suppress large-scale
fluctuation of ${\it past }$ material history.  (Historical cosmology
distinguishes ``present'' from ``past'', even though such a distinction
has played no role in this paper.)  ``Many worlds''${}^{(7)}$ i.e.,
material past histories that are distinct at scales far above
particle scale -- are not anticipated.

We conclude with superficial remarks about the notion that ``a photon
couples to electric charge.''  The general ``pair of pants'' pattern
joins 3 towers with continuity of history-chain
segments\footnote {
{\it Some} history-chain segments must
{\it reverse} their direction of age change in any ``pair of pants''
    pattern (a requirement that excludes the sector indices
 $T_1,\bar{T}_1, T_2, \bar{T}_2$ from material history); the sign of 
${\it physical}$ charge
    carried by the reversing segment corresponding reverses, just as in a
    Feynman graph.}.  ~It is
plausible that a photon ring quartet carries electric charges +e,
${\it -}$e, 0, 0, with e one-third the magnitude of charge carried by elementary
electrons.  The ring quartet of an elementary particle capable of
emitting or absorbing a photon must then include at least one
pre-event of zero electric charge and at least one of charge $\pm$e. Total electric-charge possibilities for charged
  elementary particles would then be ${\pm e, \pm 2e, \pm 3e}$.

\vfill

\noindent
{\bf Acknowledgements}
\vskip .3cm

Suggestions from J. Finkelstein and H. P. Stapp have importantly
influenced this paper, as have certain ideas of Whitehead
$^{(8)}$ and of Schwinger$^{(9)}$.

\vskip 1.2cm

\noindent
\centerline{\bf References}
\begin{enumerate}

\item G.F.~Chew, {\it Quantum Cosmology on a 2-time, Multiscale
    Coherent-state Dyonic-History Lattice,} Berkeley Lab preprint,
  LBNL-42648, to be submitted to Phys. Rev. D.

\item G.F.~Chew,  Pre-Event Magneto Electrodynamics, Berkeley Lab
  preprint, LBNL-42647, to be submitted to Phys. Rev. Letters.

\item E.W. Kolb and M.S. Turner, {\it The Early Universe},
  Addison-Wesley, 
New York~(1990).

\item P.A.M.~Dirac, Phys~ Rev. {\bf 74}, 817~(1948).

\item E.A.~Milne, {\it Relativity, Gravitation and World Structure}, Clarendon
Press, Oxford (1935).

\item S. Coleman and S. Glashow, Phys. Lett. B, {\bf 405}, 249~(1997).

\item H. Everett, Rev. Mod. Phys. {\bf 29}, 454~(1957).

\item A.N. Whitehead, {\it Process and Reality,} MacMillan, New York~(1929).

\item J.~Schwinger, Phys.~Rev.~{\bf 173}, 1536~(1968); Science {\bf 165},
757~(1969).

\end{enumerate}

\end{document}